\documentclass[11pt]{article}
\usepackage[]{graphicx}
\usepackage[]{epsfig}

\newcommand{\be}{\begin{equation}}
\newcommand{\bea}{\begin{eqnarray} \nonumber}
\newcommand{\ee}{\end{equation}}
\newcommand{\eea}{\end{eqnarray}}

 \def\(({\left(}
 \def\)){\right)}
\def\[[{\left[}
\def\]]{\right]}

\def \form#1 {eq. (\ref{#1}) }
\def \parziale#1#2  {{\partial {#1} \over \partial {#2}}}

\def \cN{{\cal N}}

\def \cB{{\cal B}}

\def \ba#1 {\overline{#1}}

\begin{document}
\title{A backtracking survey propagation algorithm for $K$-satisfiability}
\author{ Giorgio Parisi \\
  Dipartimento di Fisica, INFM, SMC and INFN, \\
Universit\`a di Roma {\em La Sapienza}, P. A. Moro 2, 00185 Rome, Italy. }
\maketitle
\begin{abstract}

In this paper we present a backtracking version of the survey propagation algorithm.  We show that
the introduction of the simplest form of backtracking greatly improves the ability of the original
survey propagation algorithm in solving difficult random problems near the sat-unsat transition.
 \end{abstract}
 
\section{Introduction}
Recently a new heuristic algorithm for solving constrain-satisfaction problems has been proposed
\cite{MPZ,MZ}.  The algorithm is based on survey propagation and it has been successfully applied to
the K-sat problem \cite{MPZ,MZ} and to the coloring problem \cite{MPWZ}.  The algorithm is well
performing in the random instance case, where the factor graph \cite{factor} associated to an
instance of the problem is locally cycle-less (i.e. it has the structure of a tree).

The properties of this survey propagation algorithm have been studied in details in the case of the
random 3-sat problem \cite{COOK} in the limit of very large number ($N$) of literals
\cite{BMZ,BMWZ,P3}.  Here the relevant parameter is
\be 
\alpha=\frac{M}{N} \ ,
\ee
where $M$ is the total number of clauses.  

The $0-1$ law \cite{KS,sat0,01} should be valid: for 
$\alpha<\alpha_{c}$ all random systems (with probability one when $N$ goes to infinity) are satisfiable.  
On the other hand, for $\alpha>\alpha_{c}$ no large random system (i.e. with probability one when $N \to 
\infty$) is satisfiable.  An 
heuristic analysis \cite{MPZ,MZ} suggests that $\alpha_{c}=\alpha ^{*}\approx 4.267$ where 
$\alpha^{*}$ can be analytically computed using the survey-propagation equations defined later.

Numerical experiments \cite{P3} shows that for very large $N$, e.g. $N=10^{6}$, the survey
propagation algorithm works up to $\alpha_{S}\approx 4.252$, that is definite less, but very near to
$\alpha_{c}$ with a time that is approximatively linear in $N$.  In other words in the region
$\alpha_{S}<\alpha<\alpha_{c}$ the problem is satisfiable for large $N$, but we are unable to find a
solution to the problem using the survey decimation algorithm.

The aim of this note is to present a backtracking version of the survey propagation algorithm that is 
somewhat slower, but it able to solve problems  where the original survey propagation algorithm does 
not work. In section II we present some general considerations on the backtracking strategies in 
the context of survey propagation. In section III we recall the basic definitions and results of the 
survey  propagation approach for random instances. Finally in the last section we present the 
details of a simple implementation of backtracking.

\section{The basic approach}

The basic idea behind survey propagation is rather simple.  In a satisfiability problem there are $L$
boolean variable (or literals) $\sigma(i)$ ($i=1,L$) that may be true or false.  We have to assign a
truth value to these literals in such a way that a given formula (i.e. the sat-formula) is satisfied.  

If we assign a value to one literal (e.g. the one indexed by $k$) we remain with a problem with
$L-1$ literals.  If we make all the correct assignments, we  start from a problem with $N$
literals and recursively we reduce  it to a null problem.  Of course making the correct assignments is a
non-trivial task.  An heuristic approach is the following.

Let us consider a given problem with $L$ variables and let us call $\cN$ is the total number of different 
truth value assignments that make the sat formula satisfied.  In a similar vein we could define 
$\cN(i)_{T}$ and $\cN(i)_{F}$ the total number of different truth value assignment of the remaining $L-1$ 
literals that make the sat-formula satisfied after that if we assign respectively a true value or false 
value to the $i^{th}$ literal.  We can thus define
\be
P_{T}(i)={\cN(i)_{T}\over \cN} \  , \ \ \ \ P_{F}(i)={\cN(i)_{F}\over \cN} \ .\label{DEF}
\ee
It may be useful to define also
\be
P(i)=\max \left(P_{T}(i), P_{F}(i)\right) \ .
\ee

Let us suppose to we know the quantities $P_{T}(i)$ and $P_{F}(i)$.  If we recursively decrease the
size of the problem from $N$ to 0 by assigning at each step the literal having the maximal value of
$P(i)$, we are going to find a correct assignment, if it exist.  While it is out of question to
compute exactly these two functions, it is possible to get some reasonable estimates of them.  The
survey equations (that will be described in the next section) provides an heuristic and efficient
way to estimate these two functions.

Generally speaking, giving an heuristic for estimating the functions $P_{T}(i)$ and $P_{F}(i)$ it is
possible to use this information to recursively decrease the size of the systems.  This is the
essence of the survey propagation method.  Indeed in many cases the survey propagation equations tell us an
estimate of the probabilities $P_{T}(i)$ and $P_{F}(i)$ that  is  the best that can be
obtained by a local analysis (i.e. an analysis such that for given $i$ $P_{T}(i)$ can be computed
with a fixed error in a time that is of order 1).

Unfortunately a local analysis is not exact and the survey estimations obtained in this way may be
not fully correct: it is possible that choices done at the initial stage turn out to be wrong.  This
may be checked by computing the increase in the number of configurations that satisfy the sat
formula if at the step $L$ we leave free the $k^{th}$ literal that was blocked at the step $M$. 
This quantity will be denoted by $1/I(k)$.  If the literal $k$ has been frozen at the previous step
(i.e. $M=L-1$), we have that 
\be P(k)=I(k) \ .  \ee 
If after freezing the other literals, when the difference $L-M$ becomes large, $I(k)$ decreases,
this is a signal that a {\sl wrong} choice has been done at the step$M$ and we would like to
backtrack this choice.  An algorithm, like survey propagation, that is able to compute the
quantities $P(k)$ is usually able to compute also the quantities $I(k)$.

These observations suggest a more general algorithm than pure decimation.  In our problem we may have 
$N-L$ literals frozen (i.e their value has been assigned) and $L$ literals free.  Our aim is to start 
from $L=N$ and to arrive to $L=0$, with the highest possible value of $\cN$ at the final destination.  We 
can consider two kind of moves: the usual decimation moves ($L \to L -1$) and the backtracking moves $L 
\to L+1$.  It seems natural in this approach to select for a decimation move the variable with largest 
$P(i)$, where $i$ labels a variable that is not assigned, and to select for a backtracking move the 
variable with smallest $I(k)$.  When
\be
P_{M}\equiv\max_{i}P(i)>I_{m}\equiv\min_{k}I(k) \ ,
\ee
a combined move, where we assign a value to the variable at $i_{M}$ (i.e. $P(i_{M})=P_{M}$) and we
deassign a value to the variable at $k_{m}$ ($I(k_{m})=I_{m}$), leads to a new configuration with
the same value of $L$ but with an higher value of $\cN$ (at least in the generic case where the
points $i_{M}$ and $k_{m}$ are far away on the factor graph and the corresponding literals are not
correlated).

\section{A survey of the survey propagation equations}

The survey propagation equations are described in details in the literature \cite{MPZ,MZ,P1,P2}.  Here we 
present only the basic ideas.

For a given problem it is convenient to consider the set of all the solutions of the problem, i.e.
of all the assignments that satisfy the sat formula.  We are we are interested not only to know the
number of all the solutions, but also to know other properties of this set.

At this end it is convenient to say that two solutions are adjacent if their Hamming distance is 
less than $\epsilon N$, where $\epsilon$ is  an appropriate small number.

It has been suggested that  in the limit of large $N$:
\begin{enumerate} 
    \item In the interval $\alpha<\alpha_{d}$ ($\alpha_{d} \approx 3.86 $ for the 3-sat \cite{P3}) the set of all 
    solutions is connected, i.e. there is a path of mutually adjacent solutions that 
    joins any two solutions of the set.  

	 \item In the interval $\alpha_{d}<\alpha<\alpha_{c}$ ($\alpha_{c} \approx 3.92 $ for the 3-sat
	 \cite{MPZ}) the set of all the solutions breaks in an large number of different disconnected
	 regions that are called with many different names in the literature
	 \cite{MP1,MP2,PARISILH,TALE,CDMM,DuMa,P3} (states, valleys, clusters, lumps\ldots).  Roughly
	 speaking the set of all the solutions can be naturally decomposed into clusters of proximate
	 solutions, while solutions belonging to different clusters are not close.
     
	 \item Only in the interval $\alpha_{b}<\alpha<\alpha_{c}$ ($\alpha_{b} \approx 3.92 $ for the
	 3-sat; in other models $\alpha_{b}$ and $\alpha_{d}$ may coincide) there are literals $\sigma$
	 that take the same value in all the legal configurations of a region.  We say that these
	 literals form the backbone of a region.  It is important to realize that a given literal may
	 simultaneously belong to the backbone of one region and not belong to the backbone of an other
	 region.

	\item Only the interval $\alpha_{m}<\alpha<\alpha_{c}$ ($\alpha_{m} \approx 4.153$ for the 3-sat
	model) the simple survey approach is correct \cite{MPR}.  For $\alpha_{b}<\alpha<\alpha_{m}$ the
	simple survey approach is likely to be an approximation and a more refined analysis is needed in
	order to get exact results.
\end{enumerate}

The last region is the hardest from the computational point of view.  Our analysis is done in this
region and we will concentrate our efforts on the properties of the backbone.  Only in the third
region the survey propagation equations have a non-trivial solution, i.e. with surveys that do not
have $s_{I}(i)$ equal to one for all $i$.  The appearance of a non-trivial solution of the survey
equation marks the entrance in the third region.

The main quantity we will study is $\cB$, i.e. the number of regions with different backbones.  As far as 
for large $N$ there is an exponential large number of configuration in each region, $\cB$ is much 
smaller than $\cN$.  One 
usually introduce the entropy \cite{MoZ} $S$ and the complexity \cite{MP2} $\Sigma$ defined as
\be
\cN=\exp(S) \ , \ \ \ \ \cB=\exp(\Sigma) \ .
\ee

In the rest of this note we will consider $\cB$ and not $\cN$ for two reason:
\begin{itemize}
    \item It is much simpler to compute $\cB$ that $\cN$. 
    \item It can be argued that the quantity $\cB$ is more interesting than $\cN$ because in the limit 
    $\alpha \to \alpha_{c}^{-}$, $\cB$ is a quantity of order 1, while $\cN$ is exponentially large ({\sl 
    nolo sumere, nondum acerbum est}). In other words the entropy density ($S/N$) does not vanish in the limit, 
    while the complexity density ($\Sigma/N$) vanishes.
\end{itemize}
In a similar way the functions $P_{T}(i)$ and $P_{F}(i)$ will be defined as the ratio of the
appropriate quantities $\cB$ (we must substitute $\cN$ with $\cB$ in eq.  (\ref{DEF})).  One hopes that
this does not make a great difference.

In the survey approach our aim is to compute $\cB$ and the surveys $\vec{s}(i)$ where $\vec{s}$ is a 
three component vector $\vec{s}= \{ s_T,s_I,s_F\}$ (the sum of the components is one).  The quantity 
$s_T(i)$ is the fraction of backbones that contain the literal at $i$ with value true, $s_F(i)$ is the 
fraction of backbones that contains the literal at $i$ with value false $s_I(i)$ is the fraction of 
backbones that do not contain the literal at $i$. If for any $i$ $s_I(i)=1$ backbones are not 
present and the problem is {\sl easy}.

From the point of view or our paper the surveys are extremely interesting because with very good 
approximation we have that
\be
P_{T}(i)=1-s_F(i)\ , \ \ \ \ \ P_{F}(i)=1-s_T(i)\  \label{MAGIC}
\ee
More precisely it can be shown that $P_{T}(i)=1-s_F(i)+O((1-s_F(i))^{2})$ that is good enough for us 
because  in most of the cases we are interested to those values of $i$ such that $P_{T}$ (or 
$P_{F}$) is very near to 1.

Unfortunately it is not possible to write local equations directly for the surveys, but we have to 
introduce the cavity surveys \cite{TAP,MPV,MP1,MP2}, i.e. the value that the survey at $i$ would have if 
we remove from the problem the clause $c$ (under the conditions that the literal at $i$ enters in the 
clause $c$).  On large random problem the cavity surveys $\vec{s}(i,c)$ satisfy simple equations that are 
called the survey propagation equations \cite{MPZ}.

The solution of these equations may be used to compute both $\cB$ and the surveys $\vec{s}(i)$. 

\section{The survey algorithm}

\subsection{Survey decimation algorithm}
The survey decimation algorithm has been proposed \cite{MPZ,MZ,BMWZ,BMZ,P3} for finding the solution of 
the random K-satisfiability problem \cite{KS,sat0}.

We start by solving the survey propagation equation and to estimate the function $P(i)$.  The main
step in the decimation procedure consists in starting from a problem with $L$ variables and in
arriving to a problem with $L-1$ variables where $\vec{s}(i)$ is fixed to be true (or false), i.e.
the literal at $i$ has an assigned value.  We have already remarked that if $P(i)$ is near to 1, the
second problem it is easier to solve: it has nearly the same number of solutions of the belief
equations and one variable less.  (We assume that the complexity can be reliable computed by solving
the survey propagation equations).

The decimation algorithm proceeds as follows.  We reduces by one the number of variables choosing the 
node $i$ in the appropriate way, e.g. by select the literal with maximal $P(i)$.  In a slightly less 
efficient but faster version  we select $R$ literals with maximal $P$, where $R=fN$ and $f$ is a small 
number.  If we are not too near the critical point and $f<.01$ the results are only weakly dependent on 
$f$.  

We recompute the solutions of the survey equations and we reduce again the number of variables; it is 
convenient to monitor the complexity.  At 
the end of the day different things may happen:
\begin{itemize}
\item We arrive to a 
negative complexity (in this case the reduced problem should have no solutions and we are lost).

\item  If in spite of this warning we go on with the algorithm,
we usually find that the reduced problem contains a contradiction  
and  in this case we are definitively lost. Sometimes we can find a contradiction also for positive 
complexity, but this happens always in the region of very small complexity density.

\item Other troubles arise when the survey equations do not have anymore an unique solution and the 
usual iterative procedure for finding the solution of the survey equations does not converge.  It is 
possible that one could find some way to escape from the troubles, but this case has never been studied 
in details.

\item After a sufficient large number of iterations the survey equation do not have any more a non-trivial 
solution and the standard iteration algorithm converges to the trivial solution ($s_{T}(i)=s_{F}(i)=0$, 
$s_{I}(i)=1$).  If this 
happens the reduced problem is  easy and it can be  solved using other algorithms.

\item An other possible happy end is that the survey equations converge to a zero entropy solution where 
(depending on the site) or $s_{T}(i)=1$ or $s_{F}(i)=1$.  In this case we do not need other algorithms to 
solve the reduced problem. 
\end{itemize}

A careful analysis  \cite{P3}  of the results for large, but finite $N=O(10^{6})$, shows that 
the algorithm works in the limit of infinite $N$ up to $\alpha_{S}(f)$ where in the limit of $f \to 0$ 
we have $\alpha_{S} \approx 4.252$, that is definite less, 
but very near to $\alpha_{c}$.

It is interesting to note that for $\alpha<\alpha_{S}$ the survey decimation algorithm takes a time
that is polynomial in $N$ and using a smart implementation the time is nearly linear in $N$.  The
survey algorithm is an incomplete search procedure that may not be able to find a solution to a
satisfiable instance.  This actually happens with a non-negligible probability e.g. for sizes of the
order of a few thousands of variables also when $\alpha<\alpha_{S}$, however in the limit $N \to
\infty$, it should work as soon $\alpha<\alpha_{S}$.  In the interval $\alpha_{S}<\alpha<\alpha_{c}$
the decimation procedure leads to a regime of negative complexity so that the algorithm does not
work.

Unfortunately at the present moment the value of $\alpha_{S}$ can be obtained only analyzing how the 
argument works on a finite sample and we are unable to compute analytically $\alpha_{S}$: it is a 
rather difficult task to understand in details why for $\alpha_{S}$ it so near to $\alpha_{c}$.

 \begin{figure}
\begin{center}
\includegraphics[width=0.65\textwidth]{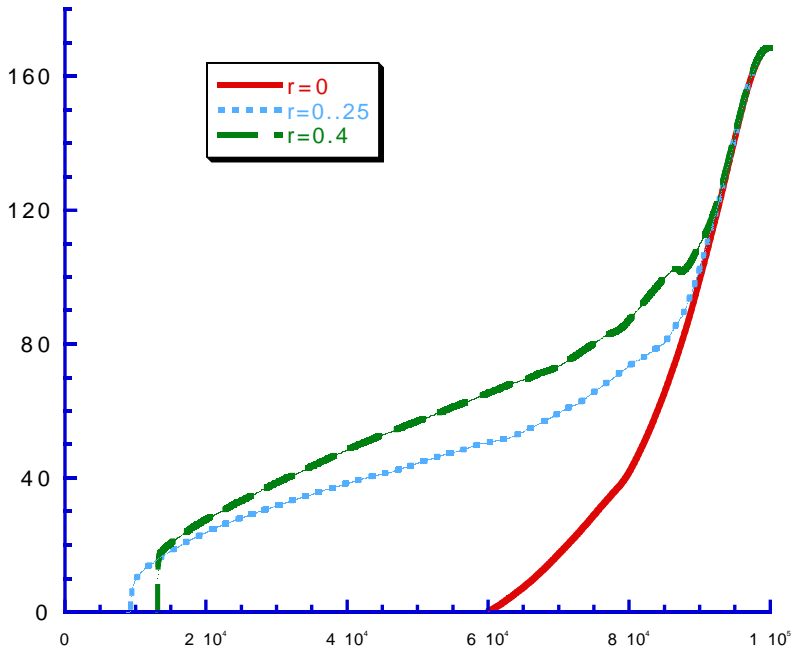}
\end{center}
\caption{The results for the $\Sigma$ as function of $L$ for one 3-sat problem with $N=10^{5}$ 
and $\alpha=4.25$ for the algorithm with $f=10^{-3}$ and three values of $r$ (0, .25 and .4,
respectively full line, short dashed line and long dashed line).}
\label{conFig}
\end{figure}

\subsection{Backtracking surveys}

In the backtracking version of the survey propagation algorithm we follow the general strategy outlined in 
the introduction.
  
In order to solve a sat problem with $N$ literals we consider auxiliary problems where $L$ literals are 
free and $N-L$ literals have a value assigned. Obviously for given $L$ we have $2^{N-L}$ of such problems.

Our aim is to start from $L=0$ and to arrive to to $L=N$, with an assignment to the literals such that 
the sat formula is satisfied.  An heuristic argument suggests that a reasonable strategy is to keep the 
highest possible value of $\cN$ at each intermediate step.

We can consider two kind of moves: the usual decimation moves ($L \to L +1$ ) and the backtracking
moves ($L \to L-1$).  In the simplest version of the the algorithm we select for applying the
decimation rule the literal with largest $P(i)$, where $i$ labels a literal that is not assigned,
and we select for a backtracking move the literal with smallest $I(k)$.  The computation of $P(i)$
is standard and $I(k)$ can be just computed in the same way as function of the surveys of the nodes
adjacent to the $k^{th}$ node.

Once we know how to select the literals to decimate and the literals to backtrack, the most important 
decision is if we do a decimation or a backtracking move.  

The simplest possibility is do decimations moves and backtracking moves at a fixed ratio.  This
algorithm depends on the parameter $r$, i.e. the ratio between the number of backtracking moves and
the total number of moves (obviously we must have $r<0.5$).  This choice has the advantage that the
total number of moves is still bounded by a number proportional to $N$ (more precisely by
$N/(0.5-r)$).  If it works, it is very efficient.

We have done a few numerical experiments and we have found that the introduction of the backtracking 
moves with the simple strategy outline before definitely increases the range of applicability of the 
survey equation algorithm.

In fig. 1. we show the results for the $\Sigma$ as function of $L$ for a 3-sat problem with $N=10^{5}$ 
and $\alpha=4.25$ for the algorithm with $f=10^{-3}$ with three values of $r$ (0, .25 and .4). In fig.2 
we plot the quantity
\be
F(L)={\sum_{i\in \mbox{non-frozen}}|s_{T}(i)-s_{T}(i)|  \over L}
\ee
where the sum is done only on the $L$ non-frozen nodes. The vanishing of $F$ implies that the solution 
of the survey equation is trivial. 
\begin{figure}
\begin{center}
\includegraphics[width=0.65\textwidth]{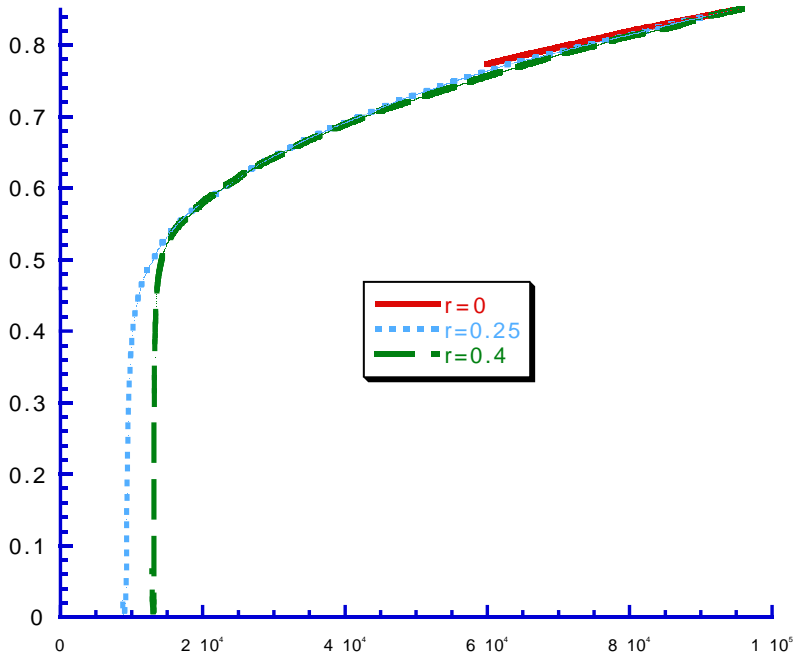}
\end{center}
\caption{The results for the the function $F$  as function of $L$ for the same problem of fig. (1) 
($N=10^{5}$, $\alpha=4.25$ and $f=10^{-3}$) with three values of $r$ (0, .25 and .4,
respectively full line, short dashed line and long dashed line).}
\label{FieldsFig}
\end{figure}

For $r=0$, i.e. the standard survey algorithm, the complexity arrives to zero with $F\ne 0$ and the
survey algorithm fails to solve the problem.  The further evolution (not shown) gives a negative
complexity and after some more iterations we find a contradiction and the algorithms must stops (as
expected).  The fact that the survey algorithm fails for a value of $\alpha$ very near, but smaller
than $\alpha_{S}$ is due to fluctuations in the value of $\alpha_{S}$ that are present for the
finite value of $N$ and vanish when $N \to \infty$.

On the contrary 
both for $r=.25$ and $r=.4$ we arrive to a situation where the complexity and $F$ simultaneously go to 
zero, i.e. to the trivial solution of the survey equations. The backtracking algorithm 
performs better that the non-backtracking one.

The backtracking algorithms is able to find an {\sl easy} reduced problem also in the case where the 
non-backtracking algorithm fails.  It is interesting to note that increasing $r$, the complexity at fixed 
$L$ increases signaling a more efficient algorithms.  Indeed there are other cases (e.g. for slightly 
larger $\alpha$) where the $r=.25$ algorithm fails and the $r=.4$ is successful.

The same phenomenon happens in a consistent way at different values of $f$ and $N$: also 
this simple version   of the backtracking algorithm is more performing that the non-backtracking 
one for difficult problems.

More carefully and systematic investigations are needed in order find out the dependence of $\alpha_{A}$ 
on the backtracking ratio $r$ and to study how near we can go to the critical point ($\alpha_{c}$) using 
this method. 

It is rather likely that the backtracking method in its present form does not work too near to $\alpha_{c}$.  Indeed 
the first decrease of the complexity (for $L$ near $N$) is weakly sensitive to the backtracking; it 
is likely using 
a survey propagation algorithm we must enter in a region of negative complexity if we are sufficiently near to 
$\alpha_{c}$.  In principle the algorithm is able to recover and to increase the complexity when we 
decrease $L$: this exactly what happens in a few cases for $N =O(10^{3})$.  

Unfortunately, when we are too 
near to $\alpha_{c}$ and we are deep in the region of negative complexity, the iterative procedure for finding 
the solution of the survey propagation equations does not converge anymore and the algorithm stop to 
work.
The non-convergence of the iterative procedure for finding the solution of the survey propagation 
equations is a sign of the existence of multiple solutions of the survey propagation equations.  It is 
likely that the next step to do is to implement a version of the algorithm that bypass this difficulty. 
It would also be very interesting to do a more careful analysis of different strategies in doing 
backtracking steps. The relative efficiency of backtracking in non-random instances is also a 
problem that one should investigate carefully.

\section*{Acknowledgements}
It is a pleasure for me to thank David Forney for a stimulating discussion.

\end{document}